\documentclass[12pt,oneside,twocolumn,english]{elsart3p}
\usepackage[latin9]{inputenc}
\usepackage{textcomp}
\usepackage{graphicx}
\usepackage{verbatim}
\usepackage{babel}
\usepackage{url}

\begin{document}

\begin{frontmatter}

\title{Performance of a Lattice Quantum Chromodynamics Kernel on the Cell Processor}

\author{J.~Spray},
\author{\corauthref{cor}J.~Hill},
\author{A.~Trew}
\address{EPCC, King's Buildings, University of Edinburgh, West Mains Road, Edinburgh, EH9 3JY}

\corauth[cor]{Corresponding author. \ead{j.hill@epcc.ed.ac.uk}}


\begin{abstract}
The implementation of a proof-of-concept Lattice Quantum Chromodynamics kernel on the Cell processor is described 
in detail, illustrating issues encountered in the porting process. The resulting code performs
up to 45GFlop/s per socket, indicating that the Cell processor is likely to be a good platform for 
future Lattice QCD calculations.
\end{abstract}

\end{frontmatter}

\section{Introduction}

The Cell processor is a heterogeneous multi-core processor\cite{Kahle05},
originally designed for high-bandwidth media applications, but also
having potential applications in traditional HPC fields such as computational science.
Interest in computational science on the Cell processor\cite{ipdps07-sweep3d,ipdps07-RAxML,Williams06}
is driven by its high peak floating point capability of up to 256GFlop/s
in single precision. Cell-based systems are available in a blade server
form factor, making possible massively parallel machines such as the
RoadRunner petaflop supercomputer currently under construction\cite{Roadrunner}.

In order to exploit the power of the Cell hardware, it is necessary
to overcome challenges associated with the programming model,
particularly the ways in which it differs from conventional processors.
The greatest difference is the Cell memory model, which replaces caches with
software-managed local stores and DMA engines.  This leads to a
requirement for explicit communication between processing cores and main memory
not found in conventional multi-core programs.  Another significant obstacle to writing 
code for the Cell or porting code to it is the SIMD instruction set used, which must 
be taken into account in order to 
obtain good floating point performance. To some
extent this difficultly may be mitigated by development tools, but
as explained in section \ref{sec:SIMD-arithmetic}, hand-optimised 
Cell-specific code is needed at present.

Since the details of the Cell architecture have been well explained in other
publications\cite{Kahle05,kistler06}, here we provide only a brief summary.
The Cell processor consists of nine cores: one Power Processing Element (PPE)
and eight Synergistic Processing Elements (SPE). The PPE is a conventional
PowerPC processor, including a 512kB L2 cache. The SPEs are SIMD processors,
optimised to operate on 128 bit vectors. Each SPE has a 256kB local
store (LS) and associated DMA engine with access to main memory as
well as the LS of other SPEs. The fabric connecting processing elements
is a bus with a peak bandwidth of 204.8GB/s, with 25.6GB/s ports for
each processing element and for the memory controller. The high bandwidth
of the bus leaves the bandwidth to the main memory controller as the
bottleneck in data movement.

The vast majority of floating point power lies in the SPEs.  In the 
best case, each of the 8 SPEs is capable of performing one 128 bit fused multiply-add operation
per cycle, corresponding to eight single precision floating point
operations, rendering a peak performance of 25.6GFlop/s per SPE on
a 3.2GHz processor.  Any code which is to perform well on the Cell 
processor must scale well enough to use all of the SPEs, and be 
written with a view to SIMD operations.  Although not all applications 
can satisfy these criteria, those based on linear algebra operations 
tend to map well to parallel systems and to SIMD instruction sets.

To illustrate the issues associated with obtaining high performance
on the Cell processor, this paper describes the process of porting
a computational kernel from Lattice Quantum Chromodynamics (Lattice QCD).
Lattice QCD simulations are extremely computationally demanding, to such
a degree that massively parallel machines have been custom-built solely
to run such calculations\cite{Kim01,Boyle03}. This paper is limited
in scope to programs on a single Cell machine, but demonstrating
the per-Cell performance possible for Lattice QCD is an important step toward
further work on a multi-node Cell system.

\section{\label{sec:Implementation}Implementation}

Prior to the work described here, various methods for Cell programming 
were experiented with by the authors\cite{dissertation}.  The implementation 
presented here ignores issues of ease-of-programming and maintainability, in 
favor of seeking the maximum possible performance. 

The SPE management library libspe2\cite{libspe2}, included in the Cell
SDK, provides basic threading functionality loading code onto an SPE
and executing it. All synchronisation and data movement is left up
to the programmer. This low level approach provides great flexibility,
but also demands relatively complex code to manage the SPEs, as well
as complex code in the SPE programs to perform DMA operations and
synchronisation as well as the actual task underway.

Creating a SPE thread using libspe2 requires only a few straightforward
function calls. Firstly, a \emph{context} is created (\texttt{spe\_context\_create}),
where a context represents one physical SPE. Next, a SPE binary is
loaded (\texttt{spe\_program\_load}) to the context, and finally the SPE program
is run (\texttt{spe\_context\_run}). The SPE program is compiled with a separate 
toolchain to the PPE program, and then embedded in the PPE executable.

Although it is possible to implement Cell-accelerated versions of 
each stage in the linear algebra procedure, and call them sequentially 
from existing procedures, this approach cannot yield good performance, since 
the bandwidth to main memory is vastly outstripped by the combined 
floating point power of the SPEs.  To make the most of the floating point 
power of the SPEs, it is necessary to persistently distribute the dataset 
across the SPEs.

\subsection{Wilson-Dslash}

Lattice QCD is a discretised form of a field theory, calculated on a 
4D space-time lattice.  Here, we refer to the spatial size of the 
lattice as $L$ and the size in time as $T$, such that the lattice 
has $L^3T$ sites in total.

A common benchmark which forms the most computationally intensive part of lattice QCD calculations
is the the inversion of the Dirac operator, a procedure known as \emph{Dslash}
\cite{gb110,WettigTilo05}.  In this paper, the physical or
mathematical meaning of these operations is not discussed: all that
is done is to directly transform an existing scalar Dslash procedure
into a Cell-optimised form.  A special-case Dslash routine is
presented which operates only in single precision, only on 4 dimensional
systems, and only on lattice data laid out in a red-black
checkerboard.  We base our Cell-optimised Dslash on a reference 
implementation provided in the freely available QDP++ package\cite{Edwards05}.
In this reference implementation, operations between neighbouring lattice sites 
are represented by operations between relatively shifted instances 
of the lattice data.  To write the Cell version of the code, all 
these operations are unrolled into explicit calculations and DMA 
operations.

% use the symbols Psi, Chi?
% is half-spinor the correct term?
% how many half-spinors do we store?

% excuse me?  What about U?
% needs defining here since it's used later when
% we discuss whether it's a constant

The Dslash procedure takes as input a \texttt{spinor} Psi 
(a $4 \times 3$ complex array at each lattice site), and four \texttt{color matrices} $u$
($3 \times 3$ complex arrays at each lattice site), one for each 
dimension. The output is the spinor Chi.  Intermediate storage is 
needed for \texttt{half-spinors} (a $2 \times 3$ complex array at each 
lattice site), which represent the so-called \texttt{projected} 
form of Psi.

\subsection{Data arrangement}

A one-dimensional block decomposition over the SPEs (of which we have $N_\mathrm{SPE}$ in total)
in the time dimension 
is used to divide the lattice into a $T/N_\mathrm{SPE}$ thick slice stored 
in the LS of each SPE.  This is inherently load-balanced as long as $T$ 
is a multiple of the number of SPEs. This is not a problematic requirement, since
existing lattice QCD simulations frequently use power-of-two sizes.
For convenience, let $T_\mathrm{chunk}=T/N_\mathrm{SPE}$. 

The operands to the Dslash operation are contiguous in main memory when using
the default QDP++ red-black checkerboard indexing scheme, wherein the slowest varying index
is the checkerboard (0 or 1), and the next slowest varying index is the
time coordinate. The calculation of an element of Chi on the SPE requires
the surrounding Psi values. A series of index manipulations are performed
to locate these values in the transferred buffers, which does not
need to be particularly efficient, since it is done once at initialisation,
and pointers to the data required for each Chi element are stored.
This look-up table sacrifices local store space for computational efficiency and
the development time saved by not having to optimise the index manipulations.

The operands are transferred to and from the SPE local stores using the 
SPE DMA engines.  Issuing DMA instructions is a straightforward low-overhead operation, 
exposed by the SDK as C intrinsics\cite{LanguageExte...}.  The addresses 
in main memory from which operands are transferred are contained in a 
structure which is loaded by the SPE threads from an address passed at initialisation 
by the PPE thread.

A beneficial characteristic of the Dslash procedure is that it is
typically used repeatedly for the same value of the gauge field $U$,
meaning that the $U$ data need only be transferred once. As such, the
time taken to transfer $U$ is negligible when the procedure is run many
times and hence the data transferred at each Dslash call is limited
to the Psi input and Chi output.  There is some scope for hiding 
communications time by double buffering, that is performing some 
calculation before Psi is completely loaded.  This is not done 
here, but ad-hoc measurements of communications time vs. calculation 
time indicate that any gains from double buffering would not 
drastically alter the final performance result.

\subsection{Padding}

Since a color matrix (a $3 \times 3$ complex matrix) has size
72 bytes in single precision, if the first element in an array of color
matrices is aligned to a 16B boundary then the subsequent element
would be offsetby 8 bytes, (72\%16 = 8). This causes a problem for SIMD operations,
since the SPE load and store instructions only work on 16 byte boundaries.
To solve this, one may either check the alignment of a color
matrix prior to operating on it, and copy it out to an aligned location
if necessary, or one may pad each element by 8 bytes. 

Since the color matrices are loaded only once at initialisation, the
cost of padding the color matrix arrays on the SPE is negligible when
Dslash is called many times. Conversely, the cost of making aligned
copies when operating on the color matrix is incurred at each invocation
of the Dslash procedure. The padding clearly incurs a penalty in the
local store space required: an increase from 72 to 80 bytes per color
matrix, or 11\%. The color matrices constitute the majority of the
local store space used for data, so this increase is significant to
the overall local store requirement. This is only a worthwhile sacrifice
as long as the serial performance of operations on the SPE is a bottleneck:
if the procedure were communications-limited then it might be preferable
to make per-operation aligned temporaries of an unpadded array and
thus accommodate a larger sub-lattice.

\subsection{\label{sec:SIMD-arithmetic}SIMD arithmetic}

SIMD-aware code is necessary to obtain even reasonable performance on the SPE, because
SPE load/store instructions operate on 
16 byte vectors, and only at 16 byte
aligned locations in the local store. To load a vector spanning a
16 byte boundary, two vector loads are required, followed by a shuffle
operation to compose the desired vector in a register. This is vastly
slower than loading a properly aligned vector, which is accomplished
in one load instruction (6 cycles).  Similarly, to perform arithmetic operations on a
scalar value and extract a scalar result requires a series of shuffle operations
in addition to the arithmetic instruction.

SIMD operations are exposed as instrinsic functions in the SPE C++
compiler\cite{LanguageExte...}, operating on 16 byte \texttt{vector} datatypes.
These allow low-level SIMD programming without the need to write assembly
language.  To perform operations on permutations of the data other than 
the native layout, the \texttt{spu\_shuffle} function is used.  This can rearrange 
the data in two input vectors to arbitrary positions across two output 
vectors.  This is used extensively in the Dslash operation, along with 
the usual arithmetic operations and the fused multiply-add provided by the SPE's FPU).

The Dslash operation has three main stages: spin projection, 
multiplication by the color matrix, and spin reconstruction. Each spin 
projection and reconstruction
function has eight variants, corresponding to forwards and backwards
in each dimension. Although mathematically each of these variants
is simply a matrix multiplication by a different matrix, in practice
they are programmatically distinct since the matrices in question
are constant, small and sparse, so the implementation of the multiplication is
unrolled into the operations required for the non-zero elements.
The color matrix multiplication has two variants, one for multiplying
a vector by the color matrix and another for multiplying by its adjoint.
Each of these 10 operations is hand-translated from scalar code in
QDP++ to SIMD intrinsics for the SPE.

Optimisation of the SIMD code is facilitated by the IBM Assembly Visualizer\cite{asmvis},
for the Cell. This tool provides a graphical display of the
execution of compiled code on a SPE, illustrating pipeline stalls and data
dependencies. The ability to see the characteristics of the code without running
it provides a very rapid turnaround and takes much of the guesswork out of
optimisation. To obtain a more detailed view of execution, the IBM Full System
Simulator\cite{mambo} may be used to model the SPE with cycle accuracy.  The simulator is also
useful for fine-grained benchmarking of SPE code since the simulated execution
environment provides convenient access to cycle counters.

As an example of the optimisation process, the forward spin projection 
phase of the calculation is considered.  Compiling the reference C
implementation in QDP++ using \texttt{spuxlc -O3 -qhot} and using
cycle counters in the Cell simulator gives a timing of 141335 cycles
to project 1024 spinors. Rewriting the arithmetic as a series of \texttt{spu\_madd}
and \texttt{spu\_shuffle} calls gives a modest runtime decrease to 98291 cycles.
Using \texttt{register vector float} type temporaries to allow the compiler to 
reorder loads and stores more aggressively reduces the timing further 
to 21084 cycles, or about 21 cycles per spinor. In the code generated
from the final version, all the multiply-add, shuffle, load and store
instructions are pipelined.

The SPE has a large register file, with 128 registers of 16 bytes each, 
easily enough to accommodate a color
matrix (5 registers), a spinor (6 registers) and many temporaries.
However, the compiler does not always take advantage of this, particularly 
it does not tend to keep values in registers between subsequent inlined functions.
To obtain the better performance this is overcome by fusing
functions which re-use the same data. Rather than having separate
projection, matrix multiplication and reconstruction functions, the
three are combined, such that all three operations can share the same
register temporaries. This requires 8 fused functions, one for each
direction. Since the color matrix multiplication functions (plain
and adjoint) are each common to 4 of the fused functions, they are
implemented as pre-processor macros, operating on register vectors floats
defined in the scope into which they are included. Although this is
substantially less elegant and flexible than having separate functions
which the compiler inlines, keeping the data in registers avoids several 
spurious loads and stores.

The SPE's fused multiply-add instruction has
the same cost as a single add or multiply, so when doing a multiply
an extra add is effectively free, and vice-versa.  This is exploited in the Dslash
procedure by implementing mixtures of subtraction and addition as multiplications
by arrays of $\pm 1$ combined with an addition.  To retain a fair comparison to 
the original code, these extra multiplications by $\pm 1$ are not counted in the 
Flop/s.  Using this scheme, our optimised spin projection function obtains a 
performance of 1864MFlop/s per SPE, compared to the original scalar code's 278MFlop/s.

The process of optimising for the Cell is utterly destructive of the original,
producing code whose operation is difficult to understand by inspection.
As such, these Cell-optimised versions must always be maintained separately
to generic code. This fits into the existing structure of QDP++, which
already includes versions of key functions optimised for the SSE architecture,
in addition to the generic implementation. 

\subsection{Extension notes}

This implementation runs only on a single cell blade. If
the Cell were to be used `in anger' for Lattice QCD then the code would have
to interleave off-node communications with on-Cell calculations and
communications. The current on-Cell parallelisation where by the division
between SPEs is in the T dimension could be retained in a multi-node
version of the code, keeping the T dimension on-node and diminishing
requirements for the inter-node interconnect to a 3D torus. If the
dataset was permanently distributed across SPEs rather than being
scattered and gathered at each iteration, it would be natural to have
the off-node communications controlled by the SPE threads, rather
than involving the PPE. In fact, given that the Cell has I/O interfaces
directly connected to the EIB, it is possible that the network interface
would be directly connected to one of these, meaning that the data-flow
could pass directly from the SPE to the network card, bypassing main
memory entirely, and ending up directly on an SPE on another node.
Whether this could be implemented effectively is a question for those
engineering the communications hardware and software stack.

In the current code, communication in the T dimension is implicit
in the distribution and collection of the fermion (Chi and Psi) data
from and to main memory at each Dslash call. However, if no further
manipulations to the data are required in between Dslash calls then
it would be more efficient to keep this data on the SPEs. In this
case, it would be necessary to implement halo swapping in the T direction
in between the SPEs. Once this is the case, the PPE would be largely
uninvolved apart from setting the SPEs running for a given
number of iterations. In this case, it may be possible to save some
amount of synchronisation runtime by implementing a peer-to-peer synchronisation
between SPEs while iterating the Dslash procedure. A higher performance
barrier may also be motivated by more frequent synchronisation, if off-node
communications demanded multiple synchronisation points during each
Dslash procedure.

Transfers larger than 16kB are currently performed by enqueuing multiple
DMA transfers of up to 16kB each. Alternatively, one could compose
a DMA list of such transfers and reduce overheads by generating this
list at startup and enqueuing it at each iteration of the algorithm.
This may yield superior performance. 

The performance of the Cell implementation is such that a Dslash over
a (6,4) sub-lattice takes 33ns per site, or 28.5\textmu{}s in total.
Using existing benchmarks of off-the-shelf interconnects\cite{cluster2004}, one
can compare this on-chip runtime to inter-node communications in a
multi-node system. Current off-the-shelf interconnects offer latencies
of the order a few \textmu{}s, while the size of the halo data for
a sub-lattice this size is of the order tens of kB, which may be expected
to incur a communications time around 10-20\textmu{}s in the best
case. This communications time is of the order
of the serial communications time, so for best sustained performance
in a multi-node system it would be imperative to overlap off-node
computation and calculation. 

Currently there is a fairly low ceiling on the size of sub-lattice
which maybe accommodated on each SPE. This limit could be made more
flexible at the cost of increased communications by streaming part
or all of the $U$ arrays at each iteration rather than requiring the
whole dataset to fit into the LS at one time. Equally, the Psi and Chi
data could be treated in a streaming manner rather than having full-sized
local arrays in the LS for the whole sub-lattice assigned to the SPE.
Some quantity of the added communications cost could be hidden using
double-buffering. This increased flexibility would be particularly
relevant in a parallel machine which may not necessarily have enough
processors for each Cell to have such as small portion of the lattice
as is currently required. 

\section{\label{sec:Performance}Performance}

All benchmarks were run on a QS20 Cell blade, with two Cell processors
and 1GB of memory for each processor.  The two processors in 
the QS20 blade are configured such that their fabrics are combined to 
give seamless access to all SPEs from one PPE program.  Benchmarks are 
run on a varying number of SPEs, denoted by $N_\mathrm{SPE}$, from 2 to 
16 (recall that each Cell processor has 8 SPEs).  Each
timing result is an average over 3 runs. Where the system size is
varied, the spatial size is denoted $L$ and the time dimension $T$,
such that there are in total $L^{3}T$ sites. As a shorthand, system
sizes are written $(L,T)$.

Performing typical strong scaling benchmarks is complicated by the
limitation on the data size which can be accommodated in the local
stores. A system size which will can be run on two SPEs is somewhat
inadequate to test the performance of 16. For this reason a variety
of system sizes were used, some providing coverage of low numbers
of SPEs, some testing larger numbers of SPEs. In addition, weak scaling
was measured using a system size with $L=2$ and $T$ proportional
to the number of SPEs used. This provides a direct test of the synchronisation
and communications, since the amount of serial work per SPE remains
constant.

\begin{figure}
\includegraphics[width=1\columnwidth]{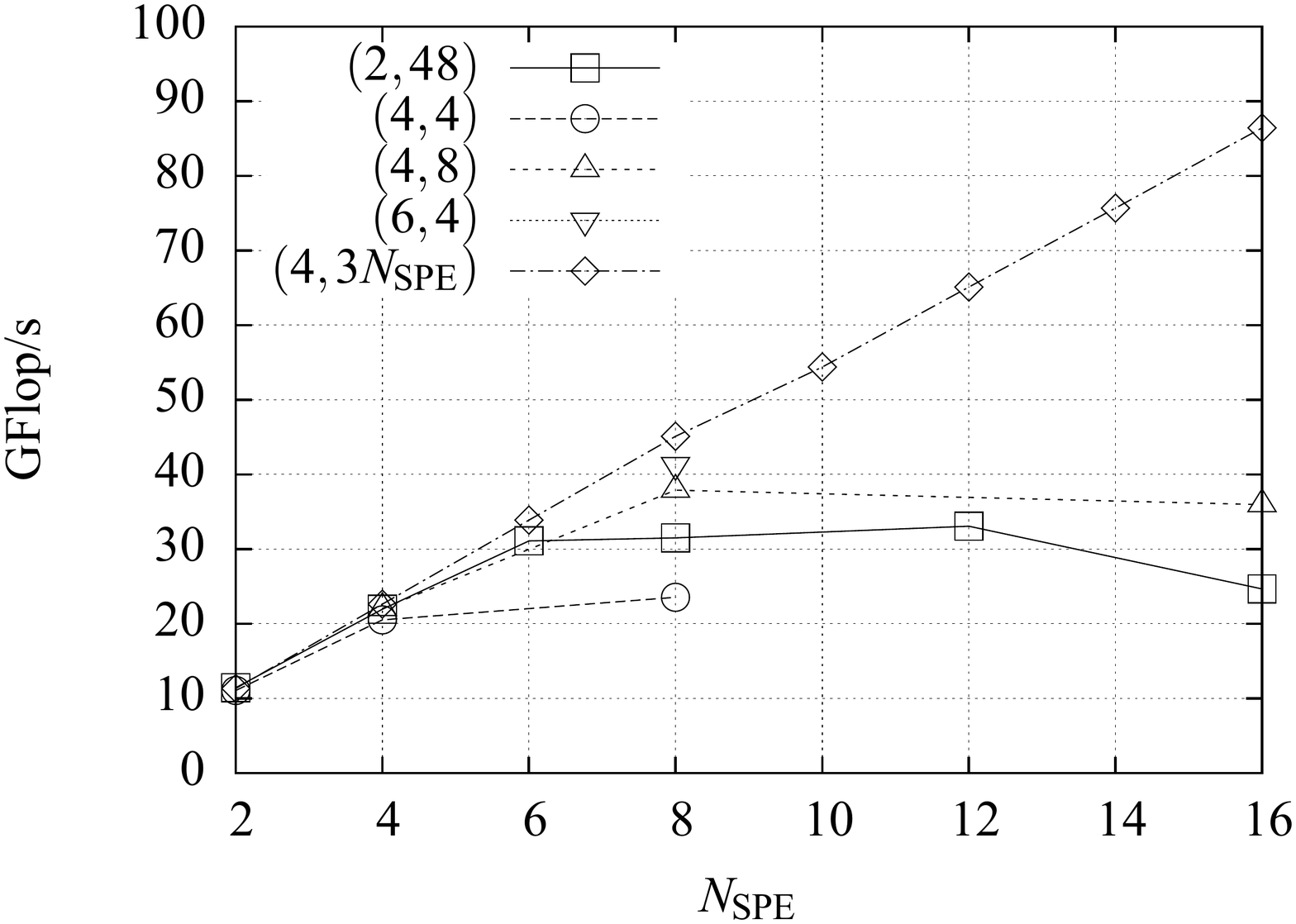}

\caption{\label{fig:Performance-of-Cell}Performance of Cell optimised Wilson-Dslash}

\end{figure}

The results of these benchmarks are shown in Fig. \ref{fig:Performance-of-Cell}.
Each series shows a different system size, including the weak scaling
case in which the system size is proportional to the number of SPEs.
In the following discussion, system sizes are expressed as (L,T) for
brevity. 

\begin{figure}
\includegraphics[width=1\columnwidth]{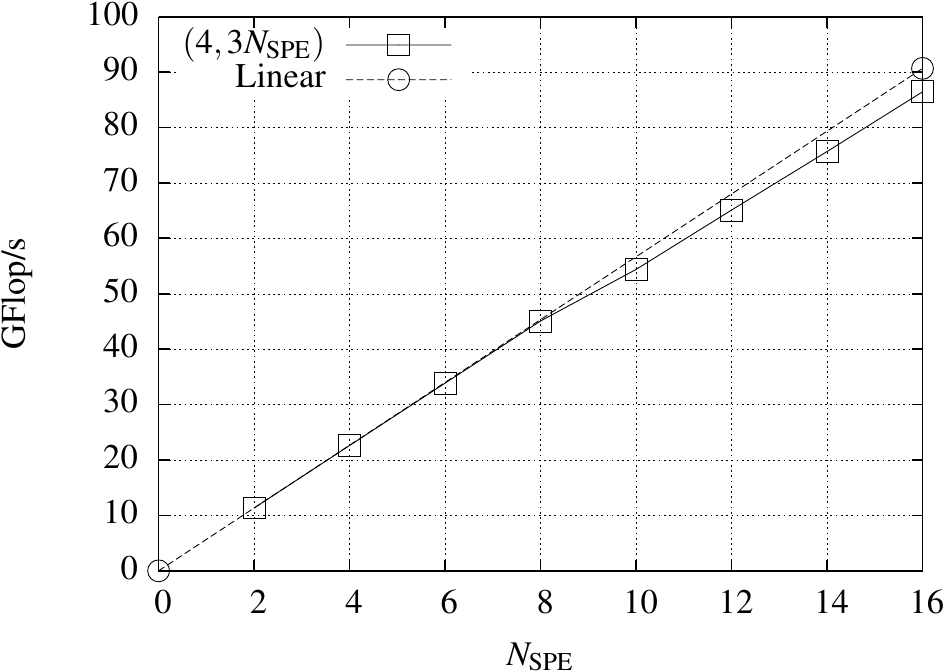}

\caption{\label{fig:linear}Weak scaling of Cell optimised Wilson-Dslash}

\end{figure}

The weak scaling case is shown in more detail in Fig. \ref{fig:linear}, where it
is compared with a straight line between the $N_{SPE}=2$ point and
the origin. There is close to linear scaling up to 8 SPEs, which falls
away for greater numbers of SPEs. This corresponds to the
regime in which some of the SPEs are on a separate chip, with a connection
between the two EIBs provided over a 25.6GB/s I/O port. The almost-linear
weak scaling plot indicates that the performance is limited largely
by serial performance. 

The $(6,4)$ system is shown in Fig. \ref{fig:Performance-of-Cell} as a single datapoint, as
it runs only on exactly 8 SPEs: a smaller number of SPEs would not
provide sufficient LS capacity, and to use more would require a larger
size in the time dimension in order to provide each SPE with at least
one 3D slice. This system is of particular interest in the context
of a multi-node system, along with the $(4,24)$ system from the weak
scaling series (the $(4,24)$ system is run on 8 SPEs). 
The $(4,24)$ system gives higher performance, but
it is quite limited in this whole-lattice time dimensions possible:
only systems with multiple-of-24 time lengths would be possible. The
$(6,4)$ system uses only around half of the available LS, but would
be much more flexible in terms of simulations on different Tsizes
across multiple nodes. A good compromise might be possible if some
quantity of the color matrix arrays were streamed from main memory
at each iteration, slightly decreasing the local store space requirement
and allowing a $(6,8)$ system on one Cell processor, at the cost
of increased main-memory I/O. 

Smaller systems which can be run on smaller numbers of SPEs are also
shown on Fig. \ref{fig:Performance-of-Cell}. These scale to smaller numbers of SPEs before performance
starts to degrade. The only system included which can be run on the
whole range of SPEs is the $(2,48)$ size, which is small enough to
run on 2 SPEs, but suffciently long in the time dimension to run on
16 SPEs. Performance of this system falls off from 6 SPEs, although
still reaches 30GFlop/s on one Cell processor (8 SPEs), which remains
impressive in comparison with conventional processors. The $(4,8)$
system achieves close to 40GFlop/s on 8 SPEs, making it a reasonably
efficient choice of system size which could be useful in multi-node
system for constructing power-of-two-sized global lattices. 

To compare cell performance to a conventional processor, a $(6,4)$
lattice is used to allow the conventional processor to work in L2
cache (this lattice is size of 145kB(1 spinor per site, 4 color matrices
per site). A larger lattice would cause the conventional platforms
to work out of main memory, while the Cell version would simply not
be capable of it. However, if the Cell version were adapted to stream
parts of the dataset out of main memory then it would probably remain
competitive with conventional platforms due to the Cell's relatively
high memory bandwidth. 

As an example of a conventional x86 processor, a 2.4GHz Xeon system was used, along with
a high performance SSE3 implementation of the Dslash operation\cite{gb110,WettigTilo05}
encapsulated in a library called intel\_sse\_wilson\_dslash. This
is included with the Chroma library and optionally compiled in on
systems which support SSE. This provides an impressive performance
of 4.5GFlop/s, demonstrating the high performance possible with this
algorithm using low-level SIMD programming. However, the single precision
floating point performance of the Cell implementation presented here
is almost 10 times greater than even the SSE-optimised version on
the Xeon.

\section{Conclusion}

The code described here obtained up to 45GFlop/s performance in single 
precision, providing a proof-of-concept for running Lattice QCD simulations
on the Cell processor.  The optimisation for Cell was accomplished entirely 
within the C environment provided by the Cell SDK, although the extensive use 
of SIMD intrinsics led to less readable code. Extensions to this work could include a multi-node parallel version
of this code, and a double precision version, which could be expected to 
achieve 20GFlop/s on the forthcoming double-precision Cell processor. However, 
the system sizes shown here would need to be revised if the Local Store on the SPUs
was not increased accordingly for a double-precision Cell.

The Cell processor remains challenging to the scientific application
programmer, demanding significant platform-specific knowledge to obtain
good performance. However, for applications concentrating work in
a single kernel -- as Lattice QCD codes do in Dslash -- the effort involved
in producing a Cell-optimised version of the kernel can render an
order-of-magnitude performance improvement over conventional platforms.

\section{Acknowledgements}

This paper was funded in part by the HPCx Terascaling project.
The authors thank IBM for providing access to Cell
hardware, in particular Joachim Jordan and Mike MacNamee. Chris Maynard and B\'{a}lint Jo\'{o}
provided useful discussion on LQCD in general and the QDP++ code in particular.

\bibliographystyle{cpc}
\bibliography{cell}

\end{document}